\documentclass[twocolumn]{aastex62}
\usepackage{graphics,epsf}
\usepackage{amsmath}                
\usepackage{amsfonts}               
\usepackage{amssymb}                
\usepackage{epsfig}                 
\usepackage{graphicx}
\usepackage{float}
\usepackage{color}
\usepackage[para,online,flushleft]{threeparttable}

\newcommand{\cm}{{~\rm cm}}
\newcommand{\km}{{~\rm km}}
\newcommand{\s}{{~\rm s}}
\newcommand{\msec}{{~\rm ms}}
\newcommand{\g}{{~\rm g}}
\newcommand{\G}{{~\rm G}}

\newcommand{\erg}{{~\rm erg}}
\newcommand{\yr}{{~\rm yr}}

\newcommand{\AU}{{~\rm AU}}

\newcommand{\days}{{~\rm days}}

\begin{document}

\title{The strongly interacting  binary scenarios of the enigmatic supernova iPTF14hls}

\author{Roni Anna Gofman}
\affiliation{Department of Physics, Technion, Haifa, 3200003, Israel; rongof@campus.technion.ac.il; soker@physics.technion.ac.il}

\author[0000-0003-0375-8987]{Noam Soker}
\affiliation{Department of Physics, Technion, Haifa, 3200003, Israel; rongof@campus.technion.ac.il; soker@physics.technion.ac.il}
\affiliation{Guangdong Technion Israel Institute of Technology, Shantou 515069, Guangdong Province, China}

\begin{abstract}
We argue that some scenarios for the enigmatic supernova (SN) iPTF14hls and its progenitor require a strong binary interaction. 
We examine scenarios that attribute the extra power of iPTF14hls to a magnetar, to a late fallback on to the neutron star (NS) that launches jets, to an interaction of the ejecta with a circumstellar matter (CSM), or to a common envelope jets SN (CEJSN).
For each of these four scenarios, we study the crucial process that supplies the extra energy and conclude that a binary companion to the progenitor must be present. 
For the magnetar scenario and late jets we claim that a companion should spin-up the pre-collapse core, in the ejecta-CSM scenario we find that the formation of the equatorial CSM requires a companion, and in the CEJSN where a NS spirals-in inside the giant envelope of the progenitor and launches jets the strong binary interaction is built-in. 
We argue that these types of strong binary interactions make the scenarios rare and explain the enigmatic nature of iPTF14hls. 
We further study processes that might accompany the binary interaction, in particular, the launching of jets before, during and after the explosion and their observational consequences. 
We do not consider the difficulties of the different scenarios and neither do we determine the best scenario for iPTF14hls. We rather focus on the binary nature of these scenarios that might as well explain other rare types of SNe.
\end{abstract}


\section{INTRODUCTION}
\label{sec:intro}

\cite{Arcavietal2017} report the discovery and evolution and  \cite{Sollermanetal2019} describe the late evolution of the extraordinary Type II supernova (SN) iPTF14hls (AT 2016bse; Gaia16aog). It is even not clear whether iPTF14hls is a core collapse supernova (CCSN), a pair instability supernova, or a common envelope jets supernova (CEJSN).  
It might turn out that iPTF14hls is not an extremely rare type of a SN, as \cite{Arcavietal2018ATel} suggest that SN 2018aad (ASASSN-18eo; \citealt{Hosseinzadehetal2018, Nichollsetal2018}) is similar in many aspects to iPTF14hls. \cite{MilisavljeviMargutti2018} in their review of peculiar SNe mention that some SNe that are initially classified as peculiar are later incorporated into the spectrum of standard events.

The peculiar properties of iPTF14hls include the following. (1) The light curve evolution is about an order of magnitude slower than that of typical type II-P SNe. (2) There are at least five peaks in the light curve. (3) There is an absorbing  circumstellar matter (CSM) with a relatively fast outflow of $v_{\rm CSM} \approx 6000 \km \s^{-1}$. For the explosion model that they assume, \cite{Arcavietal2017} find the kinetic energy of the absorbing gas to be  $E_{\rm CSM} \approx 10^{52} \erg$. However, some scenarios for iPTF14hls do not need such a large kinetic energy, e.g., \cite{AndrewsSmith2018}.  

\cite{Sollermanetal2019} estimate that iPTF14hls emitted $E_{\rm rad} = 3.6 \times 10^{50} \erg $ from discovery to their last observation. As we argue later, this energetic emission hints that iPTF14hls could not have been driven by neutrinos, as the explosion energy most likely was $E_{\rm exp} > 2 \times 10^{51} \erg$.
\cite{Sollermanetal2019} argue that the fast luminosity decline in the third year excludes the late mass accretion scenario and the magnetar scenario for iPTF14hls. They find the ejecta-CSM interaction to better fit their observations, but like \cite{Woosley2018} they find that none of the scenarios can fit all properties of iPTF14hls.  

The peculiar nature of iPTF14hls motivated several theoretical scenarios. Examples include scenarios that build on a rapidly rotating magnetic neutron star (NS), i.e., a magnetar (e.g., \citealt{Arcavietal2017, Dessart2018, Woosley2018}), some that  attribute the prolonged powering to fallback accretion (e.g., \citealt{Arcavietal2017, Wangetal2018, Liu2019}) or interaction of the ejecta with a circumstellar matter (CSM; \citealt{AndrewsSmith2018}), while others explicitly discuss a binary companion (e.g., \citealt{SokerGilkis2018iPTF14hls}). A different class of models consider the possibility that iPTF14hls is a pair instability supernova (e.g., \citealt{Woosley2018, VignaGomezetal2019}). 
 
Jets play significant roles in some scenarios, either only in the explosion itself (e.g., \citealt{Chugai2018}), or both in the explosion and a long-lasting powering of iPTF14hls, as in the CEJSN scenario \citep{SokerGilkis2018iPTF14hls} and in the scenario of late accretion of hydrogen-rich gas with stochastic angular momentum \citep{Quataertetal2019}, or only in a late powering. \cite{Liu2019} attribute the prolonged activity of iPTF14hls to the launching of jets by a black hole (BH) that continues to accrete mass for a long time, hundreds of days and more.  

As well, it might be that in all cases where magnetars supply a non-negligible amount of energy, jets play a more significant role than the magnetar in the total energy budget
\citep{Soker2016Magnetar, SokerGilkis2017Magnetar}.

\cite{Wangetal2018} propose that intermittent fallback accretion of $\approx 0.2 M_\odot$ powered iPTF14hls. In their scenario the total ejecta mass in the explosion is about $21 M_\odot$ and the explosion energy is $2.2 \times 10^{51} \erg$, that they argue can be driven by the delayed neutrino mechanism. They do not mention any binary model, and it is not clear why this explosion should be different from other CCSNe. 
They could not fit the third peak in the light curve with a fallback model, and attribute it to a magnetic outburst on the central NS.

In the present study we examine some aspects of scenarios for iPTF14hls that, we argue, require strong stellar binary interaction. We consider only cases where the explosion energy comes from gravitational energy, and we do not consider thermonuclear models. \cite{VignaGomezetal2019} claim for pair instability supernova resulting from the merger of two giant stars of about equal mass. This is another scenario for iPTF14hls of a strongly interacting binary system, but we do not consider it in the present study. 
 
We do not rank the scenarios we study by how well they fit the observations of iPTF14hls. We limit ourselves to examine the implications of the scenarios on binary interaction, as even if a scenario does not fit iPTF14hls it might fit newly discovered peculiar SNe in the future. 

In section \ref{sec:magnetar} we discuss the magnetar model that \cite{Woosley2018} proposes,  in section \ref{sec:removal} we discuss the shocked CSM scenario that \cite{AndrewsSmith2018} propose and which  \cite{MilisavljeviMargutti2018} support based on new observations (unpublished), and in section \ref{sec:Hydrogen} we briefly examine late fall back from the hydrogen rich envelope. In section \ref{sec:CEJSN} we introduce a magnetar to the CEJSN scenario. In section \ref{sec:summary} we summarise our conclusion that some of the scenarios for peculiar SNe that might account also for iPTF14hls require a strong binary interaction. 

\section{The magnetar scenario}
\label{sec:magnetar}

\subsection{The properties of the magnetar}
\label{subsec:MagnetarScenario}
As \cite{Arcavietal2017} already mentioned, one possible long-lasting power source for iPTF14hls is the rotational energy of the central NS as it spins down by dipole radiation, i.e., a magnetar. 
Let $R_{\rm NS}$ be the radius of the  magnetar,  $I_{\rm NS}$ its moment of inertia, $B$ its magnetic field, and $P$ its rotation period. The initial rotational energy of such a magnetar is $E_{\rm p} \approx 2 \times 10^{50} \left( I_{\rm NS}  / 10^{45} \g \cm^2 \right) \left( P /10 \msec \right)^{-2} \erg$. Its magnetic dipole radiation power is $P_{\rm p} \approx 4.54 \times 10^{50} (B/ 10^{14} \G)^2 (R_{\rm NS} / 12 \km)^{6} (P/10 \msec) ^{-4} \erg \yr^{-1}$, and it can power a supernova to have a maximum luminosity of (e.g., \citealt{KasenBildsten2010}) 
\begin{equation}
    L_{\mathrm{peak}} \approx f \frac{E_{\rm{p}}t_{\rm{p}}}{t_{\rm{d}}^{2}} \left[ \ln \left( 1+\frac{t_{\rm{d}}}{t_{\rm{p}}} \right) -\frac {t_{\rm{d}}}{t_{\rm{d}}+t_{\rm{p}}} \right], 
    \label{eq:Lpeak}
\end{equation}
where $f$ is a correction parameter, $t_{\rm d}$ is the photon diffusion time, and $t_{\rm p}$ is the spin-down time scale of the magnetar.  
By comparing eq. (\ref{eq:Lpeak}) to numerical simulations \cite{KasenBildsten2010} find $f=3/2$. 
The diffusion time depends on the the ejecta properties according to $t_{\rm d}=\left( 3 \kappa M_{\rm ej}/ 4\pi v_{\rm f} c \right)^{1/2}$ where $\kappa$ is the ejecta opacity, $M_{\rm ej}$ is the ejecta mass, $v_{\rm f}=\sqrt{ \left( E_{\rm p} +E_{\rm SN} \right) / 2 M_{\rm ej}}$ is the final characteristic ejecta velocity, and $E_{\rm SN}$ is the kinetic energy of the SN explosion itself. The spin-down time scale is given by 
\begin{equation} 
\begin{aligned}
    t_{\rm p} = \frac{E_{\rm p}}{P_{\rm p}} \approx & \, 0.44
    \left( \frac{P}{10 \msec} \right) ^{2} 
    \left( \frac{B}{10^{14} \G} \right) ^{-2} 
    \\ & \times
    \left( \frac{I_{\rm NS}}{10^{45} \g \cm^2} \right) 
    \left( \frac{R_{\rm NS}}{12 \km} \right)^{-4} \yr .
    \label{eq:tp}
\end{aligned}
\end{equation}

At late times after peak luminsoity the SN light curve follows the spin-down power $L_{{\rm p}}(t) =E_{{\rm p}}t_{{\rm p}}^{-1}\left(1+t/t_{{\rm p}}\right)^{-2}$. By fitting the late time light curve of iPTF14hls to $L_{\mathrm{p}}\left(t\right)$ \cite{Arcavietal2017} find that a magnetar with intial spin period of $P=5 \msec$ and an initial magnetic field of $B=0.7\times10^{14}\G $ can produce the observed average luminosity and timescale.

\cite{Dessart2018} shows that a magnetar with $P\approx 7\msec$, $I_{\rm NS}=10^{45} \g \cm^2$,  $R_{\rm NS} = 10 \km$ and $B = 0.7 \times 10^{14} \G$ that powers a typical Type II SN ejecta with $M_{\rm ej} = 13.35M_{\odot}$ and $E_{\rm SN}=1.32\times10^{51} \erg$ can produce most of the observed properties of SN iPTF14hls. Using these parameters we find that the spin-down time scale is $t_{\rm{p}}\approx 1.3 \yr$. 
\cite{Dessart2018} suggests that the ejecta is hydrogen-rich with $X\approx0.56$, therefore the opacity is mostly of electron scattering. From the opacity $\kappa_{\rm es}\approx0.3 \cm^2 \g^{-1}$ we find the diffusion time to be $t_{d}\approx0.6  \yr$. We find that these parameters give a peak luminosity of $L_{\rm peak}\approx4.35 \times 10^{42} \erg \s^{-1}$ (eq. \ref{eq:Lpeak}). We also find for these magnetar and ejecta parameters that within $450 \days$ the magnetar supplied $E_{\mathrm{mag, rad}} \approx 1.7 \times10^{50}\erg$ to radiation. This is only $\simeq 42 \%$ of the initial energy of the magnetar and is very close to the measured energy of $2.2\times10^{50}\erg$ \citep{Arcavietal2017}. We note though that the total emitted energy in the first three years is $3.6 \times 10^{50} \erg$ \citep{Sollermanetal2019}.  
   
The main point we take from this subsection is that the magnetar scenario for iPTF14hls, as well as for other peculiar CCSNe, requires a rapidly rotating newly born NS. Such a rapid rotation most likely requires the NS to  accrete mass with large angular momentum, and this accretion flow most likely launches jets (e.g., \citealt{Soker2016Magnetar, SokerGilkis2017Magnetar}).
A single star is not likely to form a rapidly rotating pre-collapse core (e.g., \citealt{Gilkis2018}), and so we suggest that the magnetar scenario for iPTF14hls requires a binary interaction with the pre-collapse core. The process is that the companion merges with the core not too long before explosion, basically long after the primary star has left the main sequence, such that the core has no time to lose its angular momentum to the envelope before explosion. Such a merger might occur in about $6-10 \%$ of all CCSNe \citep{Soker2019SCPMA}. The tidal interaction just before merger, and more so the merger itself, spins-up the core to the desired degree. Even if the merger process spins-up only the outer layers of the core, most likely the time-alternating convective zones of the core years before explosion (e.g., \citealt{Collinsetal2018, Peresetal2019}) and/or magnetic fields (e.g. \citealt{Wheeleretal2015, Peresetal2019}) transfer angular momentum to the inner layers of the core.

We also note that one of the problems of the magnetar scenario for iPTF14hls is that  it predicts a too strong emission in the third year \citep{Sollermanetal2019}. 
Nonetheless, we conduct a more general study of the magnetar scenario and hence we turn to introduce jets into the magnetar scenario.  

\subsection{A Double Phase Magnetar}
\label{subsec:MagnetarJets}
   
\cite{Woosley2018} suggests a magnetar scenario with two magnetar phases. The first magnetar phase lasts for $t_{\rm p,1} \simeq 10^{4} \s$ while the crust of the NS is forming. During this phase the NS has a high magnetic field of $B_{\rm p,1}=2\times10^{15} \G$ and a high rotational energy of $E_{\rm p,1}=2-15\times10^{51} \erg$. At the beginning of the second magnetar phase the already relaxed NS has a magnetic field of $B_{\rm p,2}=4\times10^{13} \G$ and a rotational energy of $E_{\rm p,2}\approx6\times10^{50}$.

We examine the possibility of replacing the first magnetar phase with jets activity. To do this we take the average value of the energy in the first magnetar phase in the scenario of \cite{Woosley2018}, namely $E_{\rm p,1} \simeq 8.5 \times 10^{51} \erg$, and a NS with radius of $R_{\rm NS}$ and mass of $M_{\rm NS}$. 
In case that the pre-collapse core was spun-up by a companion (section \ref{subsec:MagnetarScenario}) to the degree that the specific angular momentum at the mass coordinate $m \simeq M_{\rm NS} \simeq 1.5 M_\odot$ is larger than that of a Keplerian motion on the surface of the NS, $j_{\rm core} \ga j_{\rm Kep}=\sqrt{G M_{\rm NS} R_{\rm NS}} \simeq 1.5 \times 10^{16} \cm^2 \sec^{-1}$, then the collapsing core forms an accretion disk around the newly born NS (e.g., \citealt{Zilbermanetal2018}). Although the NS also has angular momentum, we can assume for our purposes that just as the NS forms, after the shock bounces of the newly born NS, it does not rotate rapidly.  

We further assume that this disk launches jets that powered iPTF14hls instead of the first phase magnetar in the scenario of \cite{Woosley2018}. Therefore, the jets in this phase carry an energy of $E_{\rm jets}=E_{\rm p,1} - E_{\rm p,2} \approx 8  \times 10^{51} \erg$.
The jets have a terminal specific energy, mostly kinetic energy, about equal to the specific escape energy from the NS, $G M_{\rm NS}/R_{\rm NS}$. This gives a terminal velocity of about the escape speed, such that the energy of the jets is $E_{\rm jets}= (1/2) M_{\rm jets} (2 G M_{\rm NS}/R_{\rm NS})$. Therefore, to carry an energy of $E_{\rm jets}$ the mass in the jets should be
\begin{equation}
\begin{aligned}
    M_{\rm{jets}} \simeq \frac{E_{\rm{jets}} R_{\rm{NS}}}{G M_{\rm{NS}}} \simeq & \; 0.026 
    \left(\frac{M_{\rm{NS}}}{1.4 M_\odot}\right)^{-1}
    \left(\frac{R_{\rm{NS}}}{12\km}\right)
    \\& \times
    \left(\frac{E_{\rm{jets}}}{8\times10^{51}\erg}\right) M_{\odot}.
\end{aligned}
\label{eq:Mjets}
\end{equation}
For a typical jets' outflow to accretion mass ratio of $\eta \simeq 0.1$, our estimate of the mass the NS accretes as it launches the jets is 
\begin{equation}
\begin{aligned}
    M_{\rm{acc}}  \simeq & \; 0.26  
    \left(\frac{\eta}{0.1}\right)^{-1}
    \left(\frac{M_{\rm{NS}}}{1.4 M_\odot}\right)^{-1}
     \\ & \times 
    \left(\frac{R_{\rm{NS}}}{12\km}\right)
    \left(\frac{E_{\rm{jets}}}{8\times10^{51}\erg}\right) M_{\odot}.
\end{aligned}     
\label{eq:Macc}
\end{equation}
This accreted mass implies that the NS stays as a NS and does not collapse to a black hole. 
Even for a value as low as $\eta =0.03$, for which the accreted baryonic mass is $\simeq 1 M_\odot$, the NS can grow from $\simeq 1.2 M_\odot$ to $\simeq 2.2 M_\odot$ of baryonic mass, implying a gravitational mass of $<2M_\odot$ which leaves it a NS. The accretion of angular momentum implies that the NS is rapidly rotating, as we now show, and most likely has a strong magnetic field. It is a magnetar.  

We take the specific angular momentum of the accreted mass to be a fraction $\beta$ of the Keplerian value on the NS surface 
\begin{equation}
j_{\rm{acc}}=\beta \sqrt{GM_{\rm{NS}}R_{\rm{NS}}}.
\label{eq:beta1}
\end{equation}
As jets carry angular momentum, $\beta <1$. 
Although we expect the collapsing inner core to have angular momentum, the specific angular momentum of the inner part of the collapsing core is smaller than that of the outer parts that collapse. In the approximate derivation to follow we simply take the specific angular momentum of the inner core that forms the inner part of the NS to be zero.     
Neglecting the angular momentum of the NS before accretion starts and neglecting angular momentum loss by the NS during the accretion phase, we find the angular momentum of the NS at the end of the  accretion phase to be
\begin{equation}
\begin{aligned}
    J_{\rm{NS}} & \simeq J_{\rm{acc}} = j_{\rm{acc}}M_{\rm{acc}}
        = \beta M_{\rm{acc}}\sqrt{GM_{\rm{NS}}R_{\rm{NS}}}
    \\ & \approx 
    7.7 \times10^{48} \beta
    \left(\frac{\eta}{0.1}\right)^{-1}
    \left(\frac{E_{\mathrm{jets}}}{8\times10^{51}\erg}\right)
    \\ & \times 
    \left(\frac{R_{\rm{NS}}}{12\km}\right)^{\frac{3}{2}}
    \left(\frac{M_{\mathrm{NS}}}{1.4 M_\odot} \right)^{-\frac{1}{2}} \erg \s.
\end{aligned}
\label{eq:JNS}
\end{equation}
The period of the NS as it enters the magnetar phase is
\begin{equation}
\begin{aligned}
    P & \simeq \beta^{-1} 
    \left(\frac{\alpha}{0.3}\right)
    \left(\frac{\eta}{0.1}\right)
    \left(\frac{E_{\rm{jets}}}{8\times10^{51}\erg}\right)^{-1}
    \\ & \times
    \left(\frac{R_{\rm{NS}}}{12\km}\right)^{\frac{1}{2}}
    \left(\frac{M_{\rm{NS}}}{1.4 M_\odot}\right)^{\frac{3}{2}} \msec,
\end{aligned}
\label{eq:Period}
\end{equation}
where we use  $I_{\rm{NS}}=\alpha R_{\rm{NS}}^{2}M_{\rm{NS}}$ with $\alpha\simeq0.3$ (\citealt{Raithel2016}). The rotational energy is 
\begin{equation}
\begin{aligned}
    E_{\mathrm{spin}} & \simeq 2.5 \times 10^{52} \beta^{2}
    \left(\frac{\alpha}{0.3}\right)^{-1}
    \left(\frac{E_{\mathrm{jets}}}{8\times10^{51}\erg}\right)^{2}
    \\ & \times
    \left(\frac{\eta}{0.1}\right)^{-2}
    \left(\frac{R_{\mathrm{NS}}}{12 \km}\right)
    \left(\frac{M_{\mathrm{NS}}}{1.4M_{\odot}}\right)^{-2} \erg.
\end{aligned}
\label{eq:Espin}
\end{equation}

\cite{Woosley2018} requires that the NS rotational energy at the beginning of the second magnetar phase be $E_{\mathrm{spin}}=6\times10^{50}\erg$. From that we find the value of $\beta$ (eq. \ref{eq:beta1}) in our scenario that replaces the first magnetar phase with jets activity  
\begin{equation}
\begin{aligned}
    \beta & \simeq 0.16
    \left(\frac{\alpha}{0.3}\right)^{\frac{1}{2}}
    \left(\frac{M_{\mathrm{NS}}}{1.4M_\odot}\right)
    \left(\frac{R_{\mathrm{NS}}}{12 \km}\right)^{-\frac{1}{2}}
    \\ & \times
    \left(\frac{\eta}{0.1}\right)
    \left(\frac{E_{\mathrm{jets}}}{8\times10^{51}\erg}\right)^{-1}
    \left(\frac{E_{\mathrm{spin}}}{6\times 10^{50}\erg}\right)^{\frac{1}{2}}.
\end{aligned}
\label{eq:beta}
\end{equation}
We conclude from this value of $\beta \ll 1$ that our scenario has a large margin of angular momentum. Namely, the jets can carry most of the angular momentum of the accreted mass, and we will still have the NS to rotate with the rotational energy that the scenario of \cite{Woosley2018} requires.  

Overall, we conclude that a phase of jets activity might replace the first magnetar phase in the scenario of \cite{Woosley2018} for iPTF14hls, and more generally might take place in some other peculiar superluminous CCSNe. 
We emphasise again that a single star can not have enough specific angular momentum to form an accretion disk after the collapse (e.g. \citealt{Gilkis2018,Zilbermanetal2018}). As discussed in section \ref{subsec:MagnetarScenario}, most likely a binary companion spins-up the pre-collapse core to have the required specific angular momentum to form an accretion disk. The requirement for a merging companion is one of the ingredients that make this scenario rare.

\section{The shocked CSM scenario}
\label{sec:removal}

\subsection{The properties of the CSM}
\label{subsec:CSMproperties}

\cite{AndrewsSmith2018} propose a scenario where the ejecta of a conventional supernova is shocked against a CSM disk/torus (for a simulation of this type of interaction see, e.g., \citealt{KurfurstKrticka2019}). This interaction, they propose, is the source of the energy that makes the evolution of the light curve very slow. The typical parameters that \cite{AndrewsSmith2018} use are as follows.
An explosion energy of $E_{\rm exp} \simeq 10^{51} -2 \times 10^{51} \erg$, an ejecta mass of $M_{\rm ej} \simeq 5-10 M_\odot$, an inner boundary of the CSM $R_{\rm in,CSM} \simeq 10-100 \AU$, and a CSM mass of $M_{\rm CSM} \simeq 5-10 M_\odot$. 
The interaction accelerates the CSM from a low velocity to a final velocity of $v_{\rm f, CMS} \simeq 1000 \km \s^{-1}$.  

We first take here a typical ejecta velocity of $v_{\rm ej} \approx 5000 \km \s^{-1}$, e.g., a mass of $M_{\rm ej} =8 M_\odot$ for an explosion energy of $E_{\rm exp} \simeq 2 \times 10^{51} \erg$.
As we show below a larger explosion energy is more favourable for their scenario, and we take the largest possible energy for a neutrino driven  supernova $E_{\rm exp} \simeq 2 \times 10^{51} \erg$. Neutrino driven explosion models cannot account for larger explosion energies, so a larger explosion energy requires the jet feedback explosion mechanism (e.g., \citealt{Soker2016Rev}). Later we consider the requirement for a larger energy even. 
  
Below we explore the ejecta-CSM interaction  with a CSM in a torus. However, the ejecta-CSM interaction does not depend on the exact geometry of the CSM. The requirement of the scenario of \cite{AndrewsSmith2018} is that the slow CSM does not cover the entire sphere, so that the ejecta forms a photosphere moving at a high velocity as it streams around the CSM. The most likely geometry of a slow CSM that covers part of a sphere is that of a torus, as there is a natural way to form a torus, i.e., a binary interaction. Due to the violent nature of the binary interaction that expels a large CSM mass (see below), we expect a thick torus rather than a thin one.

To account for the radiated energy of $E_{\rm rad} = 2 \times 10^{50} \erg$ \citep{Arcavietal2017}, \cite{AndrewsSmith2018} assume that the torus intercepts $10 \%$ of the ejecta. Namely, a mass of $0.8 M_\odot$ for the values we use here. From observation the CSM velocity is $v_{\rm f, CMS} \simeq 1000 \km \s^{-1}$, and so momentum conservation implies that the mass of the CSM is $M_{\rm CSM} = 3.2 M_\odot$, which together with the intercepted ejecta amounts to $4 M_\odot$. The final kinetic energy of this mass is $4 \times 10^{49} \erg$, which is about 20 per cent of the radiated energy. To account for the final kinetic energy of the CSM and ejecta, the CSM should actually intercept a larger mass, $0.125 M_{\rm ej} \simeq 1M_\odot$. As it is 4 times more massive, the CSM mass is about $4 M_\odot$. Namely, about 30 to 3 years before explosion the progenitor had lost about third of its mass in a concentrated equatorial outflow. This requires a very strong binary interaction, probably the onset of a common envelope phase. We return to this point in section \ref{subsec:CSMformation}. 

With the new estimate of \cite{Sollermanetal2019} that iPTF14hls emitted 1.8 times more energy, namely, $E_{\rm rad} = 3.6 \times 10^{50} \erg $, this CSM torus scenario becomes more extreme even. Either the explosion was more energetic with $E_{\rm exp} \simeq 3.6 \times 10^{51} \erg$, which  definitely requires a jet-driven explosion \citep{Soker2016Rev}, or the CSM intercepted a fraction of $0.125 \times 1.8=0.225 $ of the ejecta mass, or $\simeq 1.8 M_\odot$. Of course, a combination of higher energy and larger intercepted ejecta mass compared with the parameters for $E_{\rm rad} = 2 \times 10^{50} \erg$ is also possible.  

To intercept an ejecta mass fraction of $0.125-0.225$ the height of the CSM torus from the equatorial plane is $h \simeq 0.125 r - 0.23 r $. Below we consider post-shock energy that goes to accelerate the post shock gas, and find a larger value of $h$.

Not all the kinetic energy of the shocked ejecta is channelled to radiation. Let us compare the photon diffusion time in the post-shock ejecta zone with its expansion time. 
As there are no details on the ejecta mass distribution with velocity and on the CSM mass distribution with radius and angle, the derivation to follow is a very crude one.
  
Let the ejecta hit the CSM torus during a time of about 600 days during which the luminosity was high. The average total mass loss rate of the ejecta is then $\dot M_{\rm ej} \approx 8 M_\odot/(600 {\rm day}) = 1.5 \times 10^{-7} M_\odot \s^{-1}$. 
The density of the gas is $\rho_{\rm ej} \approx \dot M_{\rm ej} (4 \pi r^2 v_{\rm ej})^{-1}$. 
Due to the high velocity the post-shock zone pressure is mainly radiation pressure, and the corresponding adiabatic index is $\gamma=4/3$. The gas is then compressed in the strong shock by a factor of 7 to a density of $\rho_{\rm post} = 7 \rho_{\rm ej}$. 
 The optical depth for radiation to escape along a distance $h$ is then 
\begin{eqnarray}
\begin{aligned}
& \tau  (h) = h \kappa \rho_{\rm post} \approx 15
\left(  \frac{\dot M_{\rm ej}}{1.5 \times 10^{-7} M_\odot \s^{-1}} \right)
\left(  \frac{h/r}{0.15} \right)
\\ & \times 
\left(  \frac{v_{\rm ej}}{5000 \km \s^{-1}} \right)^{-1}
\left(  \frac{\kappa}{0.3 \cm^2 \g^{-1}} \right)
\left(  \frac{r}{10^{15} \cm} \right)^{-1} ,
\end{aligned}
\label{eq:tau}
\end{eqnarray}
where $\kappa$ is the opacity. Because of the axisymmetrical flow and the dense torus, the relevant diffusion direction is only perpendicular to the equatorial plane. This implies that the diffusion time along a distance $h$ is about 
\begin{eqnarray}
\begin{aligned}
& t_{\rm diff} (h) \simeq  \frac{ 3 h \tau(h)}{c} 
\approx 2.6 
\left(  \frac{\dot M_{\rm ej}}{1.5 \times 10^{-7} M_\odot \s^{-1}} \right)
\\ & \times 
\left(  \frac{h/r}{0.15} \right)^2 
\left(  \frac{v_{\rm ej}}{5000 \km \s^{-1}} \right)^{-1}
\left(  \frac{\kappa}{0.3 \cm^2 \g^{-1}} \right) ~{\rm day}.
\end{aligned}
\label{eq:tdiff}
\end{eqnarray}
The outflow time from that region is $t_{\rm flow}(h) \approx h /C_s$, where $C_s$ is the sound speed in the post-shock zone. For an adiabatic index of $\gamma=4/3$ this reads
$C_s=0.4 v_{\rm ej}$, and the flow time out of the post-shock region is 
\begin{eqnarray}
\begin{aligned}
t_{\rm flow} (h) \simeq & 8.7 
\left(  \frac{v_{\rm ej}}{5000 \km \s^{-1}} \right)^{-1}
\left(  \frac{h/r}{0.15} \right) 
\\ & \times
\left(  \frac{r}{10^{15} \cm} \right) ~{\rm day}.
\end{aligned}
\label{eq:tflow}
\end{eqnarray}
The fraction of the thermal energy that goes to accelerate the gas that outflows from the post-shock region instead of to radiation is $\approx t_{\rm diff} ( t_{\rm flow} + t_{\rm diff})^{-1}$.  
For the parameters we are using here this fraction is $\simeq 0.23$. This implies that to account for the radiated energy the CSM should intercept a larger fraction of the ejecta.  This new fraction is about $0.15-0.28$ instead of the fraction of $0.125-0.225$ that we derived above. 

A fraction of $0.15-0.28$ of the ejecta amounts to a kinetic energy of $3 \times 10^{50} -5.5 \times 10^{50} \erg$ for an explosion energy of $E_{\rm exp} = 2 \times 10^{51} \erg$. 
For the lower value in this range that is appropriate for a radiated energy of $E_{\rm rad} = 2 \times 10^{50} \erg$, we crudely find that a fraction of $0.2$ of the interaction energy goes to accelerate the torus in the radial direction and a fraction of $\approx 0.23$ to accelerate the shocked gas mainly perpendicular out of the interaction region. 
This implies that the energy that is channelled to radiation is only a fraction of $\approx 1-0.2-0.23=0.57$ of the energy of the gas that hits the CSM torus, or $\approx 1.7 \times 10^{50} \erg$. To reach a value of $2 \times 10^{50} \erg$ in radiation, we need to take $h=0.19 r$, keeping all other parameters the same. 

For the larger radiated energy of $E_{\rm rad} = 3.6 \times 10^{50} \erg$ that \cite{Sollermanetal2019} estimate more recently, the value is larger, crudely $h \simeq 0.3r$, or the explosion energy is larger, as we explained above. 

We note that if the inner edge of the CSM torus is at $r=0.5 \times 10^{15} \cm$ instead of $r=10^{15} \cm$, then the fraction of the energy that goes to accelerate the shocked gas is $\approx 0.37$ rather than $\approx 0.23$, keeping all other parameters as in equations (\ref{eq:tdiff}) and (\ref{eq:tflow}). Namely, a lower fraction of the energy goes to radiation hence requiring a more energetic explosion even. 
 
Our conclusion from this discussion is that for an explosion energy of $E_{\exp}=2 \times 10^{51} \erg$ the CSM should intercept a fraction of about $0.3$ of the ejecta. The CSM mass should be about equal to the ejecta mass $M_{\rm CSM} \approx M_{\rm ej}$. 
For an explosion energy of $E_{\exp}=3.6 \times 10^{51} \erg$ the CSM should intercept a fraction of about $0.15-0.2$ of the ejecta, and the CSM mass should be $M_{\rm CSM} \approx 0.6-0.8 M_{\rm ej}$. 
 
Observations do not show the CSM disk at early times, and \cite{AndrewsSmith2018} suggest that the ejecta flows around the CSM disk and hide it. 
When the shocked ejecta passes the CSM disk it expands transversely at about its sound speed. To close the photosphere behind the CSM disk of half thickness of $h \simeq (0.15-0.2)r$ the sound speed should be $\approx 0.15-0.2 v_{\rm ej}$ or larger. This implies that the thermal energy of the gas is about three per cent or more of its kinetic energy.  By the time a post-shock parcel of gas of the ejecta flows around the CSM disk it supposes to have lost most of its thermal energy to radiation, so it is not clear it will have such a high sound speed. Hydrodynamical simulations that include radiative transfer are required to explore the nature of this outflow.     

\subsection{The formation of the CSM disk}
\label{subsec:CSMformation}

The parameters we derive for the CSM in section \ref{subsec:CSMproperties} make this supernova extreme according to the shocked CSM scenario. Its explosion is on the upper side of traditionally neutrino-driven supernovae, or above that limit. As well, the ejection of a mass of $M_{\rm CSM} \approx 0.6-1 M_{\rm ej}$ at a time of $30 \yr$ (for initial CSM velocity of $10 \km \s^{-1}$) to $3 \yr$ (for initial CSM velocity of $100 \km \s^{-1}$) before explosion requires a specific explanation. 
It seems to us that the ejection of so much mass in an equatorial outflow just tens of years before explosion requires a very strong binary interaction with a relatively massive companion. Such a binary interaction can be observed as a giant eruption of a Luminous Blue Variables (LBV) or as a SN impostor (e.g. \citealt{Kashi2018}).

The question then is why did the binary interaction take place just $\approx 3-30 \yr$ before explosion? 
The mechanism to drive a strong binary interaction before explosion is an instability in the supernova progenitor that causes its envelope to substantially expand, e.g., because of vigorous convection in the core that cause waves to expand into the envelope \citep{QuataertShiode2012}, or because of magnetic activity that carries energy from the core to the envelope \citep{SokerGilkis2017}. Such energy deposition to the envelope causes its expansion (e.g., \citealt{Soker2013, Fuller2017}). If there is a close binary companion, then a strong interaction can take place (e.g., \citealt{McleySoker2014, DanieliSoker2019}). 
 
There is another possibility that also involves a massive binary stellar companion, but where the interaction took place over a long time. In this scenario a close companion by its gravity ejected some of the primary envelope mass with a large specific angular momentum, such that the mass formed a long-lived Keplerian disk around the binary system. This scenario does not have to account for the coincidence between disk/torus formation and the explosion, as the disk is long-lived. 
Such long-lived circumbinary disks are observed in some evolved low mass stars. These post asymptotic giant branch stars have a main sequence companion at an orbital separation of $\approx 1\AU$ and a circumbinary disk with radii of tens to hundreds of AU (e.g., \citealt{Kastneretal2010, VanWinckel2017a}). In some of these systems the main sequence companion   launches jets (e.g., \citealt{Wittetal2009, Gorlovaetal2012, Thomasetal2013,  VanWinckel2017b}). 

The same process where a companion ejects mass in the equatorial plane and forms a long-lived disk and at the same time it launches jets might take place with more massive binary stars that are progenitors of core collapse supernovae. \cite{Soker2017IIb} proposes that some progenitors of core collapse supernovae experience such an interaction, and if the companion manages to eject most of the hydrogen-rich envelope the outcome will be a Type IIb supernova. The companion is outside the giant envelope, but very close to the surface, in what is termed the grazing envelope evolution. 
If this scenario holds, then the ejecta of iPTF14hls collides not only with equatorial CSM but also with polar CSM. However, the formation of a so massive CSM disk requires an extreme type of interaction. 
   
There is a third possibility for such an extreme interaction, where both stars were giants during the interaction. While the primary was close to explosion, the secondary was at an earlier stage, like still having a helium-rich core, but its radius was large.
The stars reach this evolutionary point when the second star expands and tidal forces spin it up on account of orbital angular momentum, so that the two stars reduce their orbital separation, until they possibly form a common envelope. Now the two cores spirals-in inside a common envelope. For the two stars to be giants at the same time their initial masses are close to each other (e.g., \citealt{Segevetal2019}). When the two stars are close to contact just before they form a common envelope, the larger giant overflows its Roche lobe. The mass loss through the outer Lagrangian point can form an expanding torus (disk; e.g., \citealt{Pejchaetal2017}). In the case of two giant stars merging we do not expect polar outflows. There is a need for a three-dimensional hydrodynamical simulation of this process to reveal the outflow geometry (present hydrodynamical simulations of the common envelope phase include only one giant, e.g., \citealt{Chamandyetal2018, Reichardtetal2019}).
 The merger process of the two giants forms a luminous transient event \citep{Segevetal2019}, and might explain the 1954 pre-explosion outburst of the progenitor of iPTF2014hls (see \citealt{Arcavietal2017} for details on that pre-explosion outburst). 
 
\section{Late accretion of hydrogen envelope gas}
\label{sec:Hydrogen}

\cite{Quataertetal2019} build a scenario based on the jittering jets explosion mechanism (e.g., \citealt{PapishSoker2011, PapishSoker2014}) where convective fluctuations in the pre-collapse star lead to the formation of a stochastic (intermittent) accretion disk around the newly born NS or BH. Following earlier studies (e.g., \citealt{GilkisSoker2014, GilkisSoker2016}) of accretion disk formation as a result of convective fluctuations in the helium shell and inward, \cite{Quataertetal2019} show that the convective fluctuations in the hydrogen-rich shell are likely to form a stochastic disk when accreted onto the newly born BH, and that this disk launches jitterring jets that explode the star. 
The accretion from the hydrogen shell occurs also long after explosion  \citep{Gilkisetal2016, Quataertetal2019}, and hence the central NS or BH can launch the jets at late times, namely, several times the stellar dynamical times, or hundreds of days in the case of a massive giant progenitor. 
 
The question in this jittering jets explosion mechanism, which in this case operates during the accretion from the hydrogen-rich envelope, is why should the system reach a stage that the hydrogen-rich envelope is accreted? Some call it a `failed supernova'. But if it eventually explodes it cannot be a `failed CCSN' \citep{Soker2017RAA}. We accept the scenario that \cite{Gilkisetal2016} suggest where such a CCSN occurs when the pre-collapse core is rapidly rotating. The accretion disk around the newly born NS or BH maintains a constant angular momentum axis, and hence launches jets along a constant axis. These jets remove stellar mass along the two opposite polar directions, but they do not manage to eject stellar gas from near the equatorial plane, and hence operate in an inefficient jet feedback mechanism. The removal of polar gas reduces gravity, and the equatorial gas expands due to pressure gradient in the envelope. Some of this gas does not reach the escape velocity from the star, and later flows back to feed the, now, central BH that hence launches late jets. 

\cite{Chugai2018} proposes a scenario with an ejected mass of $30 M_\odot$ and an explosion energy of $8 \times 10^{51} \erg$. The explosion is driven by relativistic jets that an accretion disk around a BH launches. He does not specify the source of angular momentum, but his scenario also requires a relatively rapidly rotating pre-collapse core. 

For the pre-collapse core to rapidly rotate at this very late stellar evolution phase, as required by the above two scenarios, it must have been spun-up by a compact stellar companion that merges with the core, i.e., a fatal common envelope evolution. The system must obey three conditions for the companion to reach the core under these conditions (\citealt{Soker2019SCPMA} for more details).
(1) The ratio of companion mass to envelope mass be low to ensure that the companion does not eject the envelope before it reaches the core. The ratio should crudely be $M_{\rm env} \ga 5 M_2$. (2) The companion should have high density such that the core does not tidally destroy it far from the core. A low mass main sequence companion, and more so a white dwarf (WD) are fine. (3) The companion should enter the envelope at late stages of evolution, so that the core does not have time to transfer most of its angular momentum to the envelope 

Overall, the companion should be a main sequence star of mass $M_2 \approx 0.5 M_\odot - 0.15 M_1$ or a WD. In the case of a WD companion and a progenitor of a CCSN, the system experiences the reverse evolution, where the WD forms before the NS does \citep{SabachSoker2014}. The more massive star in a binary system has a zero age main sequence mass of $5.5 M_\odot \la M_{\rm ZAMS,2} \la 8.5 M_\odot$ \citep{Soker2019SCPMA}. It evolves first to form the WD of mass $M_2 \simeq 1 M_\odot$. It transfers mass to a companion, that after the accretion phase has a mass of $M_1 \ga 9M_\odot$. The reverse evolution requires specific properties of the binary system, making it a rare event. \cite{Soker2019SCPMA} estimates that this fatal common envelope reverse evolution occurs in about two per cent of all CCSNe. 

The core destroys a main sequence companion at a radius of $\approx 1 R_\odot$, while a WD will enter the core at the radius of the core, $R_{\rm core} \approx 0.1 R_\odot$. The angular momentum of the companion is 
$J \approx M_2 (G M_{\rm core} R_{\rm core})^{1/2} $. 
For $M_2=0.5 M_\odot$, $R_{\rm core} = 0.1 R_\odot$, $M_{\rm core} = 5 M_\odot$, the specific angular momentum of the core after merger with the companion is $j_{\rm core} \approx 2 \times 10^{17} \cm^2 \s^{-1}$. 
We compare it to the specific angular momentum of a Keplerian disk around a NS of mass $1.4 M_\odot$ and a radius of $15 \km$, $j_{\rm NS,disk} \simeq 2 \times 10^{16} \erg \s^{-1}$.
This scenario has a large margin to allow a collapsing core to form an accretion disk around the newly born NS and BH. 

We summarise this section by first stating again that the jittering jets explosion mechanism itself does not need the core to rotate at all. However, in the scenario of \cite{Quataertetal2019} where the accreted gas comes from the hydrogen-rich shell we should explain why the star did not explode earlier. For that we employ the scenario of \cite{Gilkisetal2016} where a rapidly rotating core might explode at late times. 
Therefore, the accretion of hydrogen-rich gas around a newly BH can take place when a companion merges with the core, i.e., a fatal common envelope evolution, and spins it up. 

\section{Forming a magnetar in the CEJSN scenario}
\label{sec:CEJSN}

\cite{SokerGilkis2018iPTF14hls} propose the CEJSN scenario for iPTF14hls, where a NS that spirals-in inside the envelope of a giant star and to the core, destroys the core. Along the entire evolution the NS accretes mass via an accretion disk and launches jets. The jets power the pre-explosion mass ejection and the explosion itself, and might power iPTF14hls for a long time after explosion. The CEJSN scenario is based on the formation of an accretion disk cooled by neutrino \citep{Chevalier1996} around a NS inside a common envelope (e.g., \citealt{FryerWoosley1998, ArmitageLivio2000, Soker2004, Chevalier2012}). 

We here introduce, but do not simulate, a scenario where the CEJSN leads to a shocked CSM tours scenario or to a magnetar spin-down scenario or even both.

Consider a binary system of a NS and a massive star.  The massive star has a mass of $\approx 80 M_\odot$, as in the CEJSN scenario of \cite{SokerGilkis2018iPTF14hls} for iPTF14hls, and it expands to become a cool giant of radius $\approx 50-100 R_\odot$. The orbital separation is not much larger than the giant radius. Even if the NS manages to bring the system to synchronisation, the system is unstable to the Darwin instability and the NS rapidly spirals in toward the core of the giant. 

For jets' energy that equals an explosion energy of $E_{\rm exp} \approx 4 \times 10^{51} \erg$ (see discussion in section \ref{subsec:CSMproperties} for the new radiated energy value from \citealt{Sollermanetal2019}), we find by using eq. (\ref{eq:Macc}) that $M_{\rm acc} \approx 0.13 M_\odot$. Equation (\ref{eq:Period}) shows that the period of the NS after the mass accretion is $P\approx 1.96 \beta^{-1} \msec$, where $\beta$ is defined in eq. (\ref{eq:beta1}). Moreover, the initial rotational energy of this magnetar is $E_{\rm p} \approx 6.2 \times 10^{51} \beta ^2  \erg$, using eq. (\ref{eq:Espin}). If we take the magnetar rotational energy to be as in the second magnetar phase of \cite{Woosley2018}, $E_{\rm p} = 6 \times 10^{50} \erg$, we conclude that $\beta \approx 0.31$. As we concluded in section \ref{sec:magnetar}, here too the jets might carry most of the angular momentum of the accreted mass and still the NS ends up as a magnetar with the rotational energy as required by \cite{Woosley2018}.

In addition to becoming a magnetar, the NS can eject a slow equatorial outflow as it spirals-in inside the envelope of the giant. Overall, the CEJSN might lead to a late magnetar activity and/or ejecta-CSM interaction. This adds to the energy of the jets that the NS launches. These complications that enrich the outcomes of the CEJSN  scenario require deeper studies.    

\section{SUMMARY}
\label{sec:summary}
 
We explored some scenarios from the literature to explain the enigmatic SN iPTF14hls. We summarise these scenarios in table \ref{table:ScenrioSummery} and list the relevant sections where we discuss our new suggestions and conclusions. In the first, column we give the name of the scenario and relevant references. The second column lists the process that leads to prolonged activity that ensures a slow light curve according to the scenario, The third column lists the critical ingredient that the scenario requires for the prolonged activity. In the fourth column, we give the binary interaction that we suggest makes this scenario rare, and in the fifth column, we list processes that might accompany the binary interaction. 
The last column gives the explosion energy of iPTF14hls according to the scenario. 

\begin{table*}
\scriptsize
\centering
\begin{center}
\begin{tabular}{|c|c|c|c|c|c|}
\hline 
                            & Prolonged              & Critical           &  Rare binary                                     & Possible Accompanying                          & Explosion Energy   \\
                            & Activity               & Ingredient         &  property                                        & processes                                      & for iPTF14hls 
\tabularnewline
\hline 
\hline   
Magnetar                    & Rapidly                & Rapidly rotating    & Companion spins-up                              & Accretion disk and                             & $\approx10^{51} \erg$   \\
$\rm{[Ar17][De18][Wo18]}$   & rotating NS            & pre-collapse core   & the core (\S \ref{subsec:MagnetarScenario})     & jets at explosion and                          & [De18][Wo18] \\ 
                            &                        &                     &                                                 & for few hours (\S \ref{subsec:MagnetarJets})   &           
\tabularnewline
\hline 
CSM                         & Ejecta-CSM             & Massive             & Interaction to eject                            & The companion                                  & $\ga 2 \times 10^{51} \erg $; Requires   \\ 
$\rm{[AS18]}$               & collision              & equatorial CSM      & equatorial mass (\S \ref{subsec:CSMformation})  & launches jets (\S \ref{subsec:CSMformation})   & explosion by jets (\S \ref{subsec:CSMformation}) 
 \tabularnewline
\hline 
Fallback                    & Late accretion         & Rapidly rotating    & Companion spins-up                              & Accretion disk and                             & $\approx10^{51} \erg$ \\ 
$\rm{[Ar17][Wa18][Li19]}$   & on to NS               & pre-collapse core   & the core (\S \ref{sec:Hydrogen})                & jets at explosion (\S \ref{sec:Hydrogen})      & [Wa18] 
\tabularnewline
\hline
CEJSN                       & (1) Ejecting a         & Accretion of mass    & NS spirals-in                                  & NS ejects equatorial                           &  (1) $\approx10^{52} \erg$  [SG18]   \\ 
$\rm{[SG18]}$               & massive envelope       & on to NS inside      & down into the                                  & mass and  turns to                             &  (2) $\approx 4 \times 10^{51} \erg$ (\S \ref{sec:CEJSN}) \\
                            & and/or (2) late jets   & supergiant star      & companion's core                               & a magnetar (\S \ref{sec:CEJSN})                & 
\tabularnewline
\hline 
\end{tabular}
\end{center}
\begin{tablenotes}
\textbf{Acronyms:}  NS: neutron star;  CSM: circumstellar matter; CEJSN: common envelope jets supernova;
\newline
\textbf{References:} [Ar18]: \citep{Arcavietal2017}; [AS18]: \citep{AndrewsSmith2018}; [De18]: \citep{Dessart2018}; [Li19]; \citep{Liu2019}; [SG18]: \citep{SokerGilkis2018iPTF14hls}; [Wa18]: \citep{Wangetal2018}; [Wo18]: \citep{Woosley2018}. 
\newline
\end{tablenotes}
\caption{A summery of scenarios proposed of iPTF14hls and discussed in this study.}
\label{table:ScenrioSummery}

\end{table*}

We recall that we did not try to find the best scenario or to point out difficulties with the different scenarios. We simply argue that any of the scenarios for iPTF14hls that we discussed here requires a strong binary interaction, or even a gravitational (tidal) destruction of one of the stars before the explosion. 

These binary interactions are likely to lead to the launching of jets in many cases, as we discussed in the respective sections and summarise here in table \ref{table:ScenrioSummery}. The binary interaction with its possible outcomes, e.g., strong jets and an equatorial outflow, explain the peculiarity of iPTF14hls and some other peculiar CCSNe. The binary interaction naturally leads to a dense equatorial outflow before the explosion. Jets that the companion might launch before explosion might form a bipolar CSM. Namely, two opposite polar fast outflows in addition to the equatorial outflow. 

The binary interactions that involve a common envelope evolution, including the CEJSN and cases where a companion merges with the core, might have a transient event just before or during the time the companion enters the common envelope. Such a transient event might account for the 1954 pre-explosion outburst of iPTF14hls (see \citealt{Arcavietal2017} for details on that pre-explosion outburst). 
\cite{SokerGilkis2018iPTF14hls} attributed the 1954 pre-explosion outburst to an eccentric orbit and temporary mass accretion by the NS at a periastron passage \citep{Gilkisetal2019} prior to the onset of the common envelope phase.

In short, our study strengthens the case for iPTF14hls and its progenitor to be an outcome of strong binary interaction. The general implication of our results is that the wide diversity of binary interactions ensures a rich variety of peculiar CCSNe. 

\section*{Acknowledgements}
We thank an anonymous referee for useful comments.
We thank Avishai Gilkis for useful comments. 
This research was supported by a grant from the Israel Science Foundation.

\label{lastpage}

\begin{thebibliography}{}

\bibitem[Andrews \& Smith(2017)]{AndrewsSmith2018} Andrews, J.~E., \& Smith, N.\ 2017, arXiv:1712.00514

\bibitem[Arcavi et al.(2018)]{Arcavietal2018ATel} Arcavi, I., Hiramatsu, D., Jha, S.~W., et al.\ 2018, The Astronomer's Telegram, 12135

\bibitem[Arcavi et al.(2017)]{Arcavietal2017} Arcavi, I., Howell, D.~A., Kasen, D., et al.\ 2017, \nat, 551, 210

\bibitem[Armitage \& Livio(2000)]{ArmitageLivio2000} Armitage, P.~J., \& Livio, M.\ 2000, \apj, 532, 540

\bibitem[Chamandy et al.(2018)]{Chamandyetal2018}  Chamandy, L., Frank, A., Blackman, E.~G., et al.\ 2018, \mnras, 480, 1898 

\bibitem[Chevalier(1996)]{Chevalier1996} Chevalier, R.~A.\ 1996, \apj, 459, 322 

\bibitem[Chevalier(2012)]{Chevalier2012} Chevalier, R.~A.\ 2012, \apjl, 752, L2

\bibitem[Chugai(2018)]{Chugai2018} Chugai, N.~N.\ 2018, Astronomy Letters, 44, 370

\bibitem[Collins et al.(2018)]{Collinsetal2018}  Collins, C., M{\"u}ller, B., \& Heger, A.\ 2018, \mnras, 473, 1695

\bibitem[Danieli, \& Soker(2019)]{DanieliSoker2019} Danieli, B., \& Soker, N.\ 2019, \mnras, 482, 2277.

\bibitem[Dessart(2018)]{Dessart2018} Dessart, L.\ 2018, \aap, 610, L10.

\bibitem[Fryer \& Woosley(1998)]{FryerWoosley1998} Fryer, C.~L., \& Woosley, S.~E.\ 1998, \apjl, 502, L9 

\bibitem[Fryer et al.(1999)]{Fryeretal1999} Fryer, C.~L., Woosley, S.~E., \& Hartmann, D.~H.\ 1999, \apj, 526, 152

\bibitem[Fuller(2017)]{Fuller2017} Fuller, J.\ 2017, \mnras, 470, 1642

\bibitem[Gilkis(2018)]{Gilkis2018}  Gilkis, A.\ 2018, \mnras, 474, 2419 

\bibitem[Gilkis et al.(2019)]{Gilkisetal2019} Gilkis, A., Soker, N., \& Kashi, A.\ 2019, \mnras, 482, 4233.

\bibitem[Gilkis \& Soker(2014)]{GilkisSoker2014} Gilkis, A., \& Soker, N.\ 2014, \mnras, 439, 4011

\bibitem[Gilkis \& Soker(2016)]{GilkisSoker2016} Gilkis, A., \& Soker, N.\ 2016, \apj, 827, 40

\bibitem[Gilkis et al.(2016)]{Gilkisetal2016} Gilkis, A., Soker, N., \& Papish, O.\ 2016, \apj, 826, 178 

\bibitem[Gorlova et al.(2012)]{Gorlovaetal2012} Gorlova, N., Van Winckel, H., Gielen, C., et al.\ 2012, \aap, 542, A27

\bibitem[Hosseinzadeh et al.(2018)]{Hosseinzadehetal2018} Hosseinzadeh, G., Hiramatsu, D., Arcavi, I., et al.\ 2018, Transient Name Server Classification Report, 373, 

\bibitem[Kasen, \& Bildsten(2010)]{KasenBildsten2010} Kasen, D., \& Bildsten, L.\ 2010, \apj, 717, 245.

\bibitem[Kashi(2018)]{Kashi2018} Kashi, A.\ 2018, Galaxies, 6, 82

\bibitem[Kastner et al.(2010)]{Kastneretal2010} Kastner, J.~H., Buchanan, C., Sahai, R., Forrest, W.~J., \& Sargent, B.~A.\ 2010, \aj, 139, 1993

\bibitem[Kurf{\"u}rst \& Krti{\v{c}}ka(2019)]{KurfurstKrticka2019} Kurf{\"u}rst, P., \& Krti{\v{c}}ka, J.\ 2019,
arXiv:1904.01312  

\bibitem[Liu et al.(2019)]{Liu2019} Liu, T., Song, C.-Y., Yi, T., Gu, W.-M., \& Wang, X.-F.\ 2019, arXiv:1902.03464

\bibitem[Mcley \& Soker(2014)]{McleySoker2014} Mcley, L., \& Soker, N.\ 2014, \mnras, 445, 2492

\bibitem[Milisavljevic \& Margutti(2018)]{MilisavljeviMargutti2018} Milisavljevic, D., \& Margutti, R.\ 2018, \ssr, 214, 68

\bibitem[Nicholls et al.(2018)]{Nichollsetal2018} Nicholls, B., Brimacombe, J., Kiyota, S., et al.\ 2018, The Astronomer's Telegram, 11391

\bibitem[Papish \& Soker(2011)]{PapishSoker2011} Papish, O., \& Soker, N.\ 2011, \mnras, 416, 1697. 
  
\bibitem[Papish \& Soker(2014)]{PapishSoker2014} Papish, O., \& Soker, N.\ 2014, \mnras, 438, 1027

\bibitem[Papish et al.(2015b)]{Papishetal2015} Papish, O., Soker, N., \& Bukay, I.\ 2015b, \mnras, 449, 288

\bibitem[Paxton et al.(2015)]{Paxton2015} Paxton, B., Marchant, P., Schwab, J., et al.\ 2015, \apjs, 220, 15

\bibitem[Pejcha et al.(2017)]{Pejchaetal2017} Pejcha O., Metzger B.~D., Tyles J.~G., Tomida K., 2017, ApJ, 850, 59

\bibitem[Peres et al.(2019)]{Peresetal2019} Peres, I., Sabach, E., \& Soker, N.\ 2019, \mnras, 486, 1652

\bibitem[Quataert et al.(2019)]{Quataertetal2019} Quataert, E., Lecoanet, D., \& Coughlin, E.~R.\ 2019, arXiv:1811.12427

\bibitem[Raithel et al.(2016)]{Raithel2016} Raithel, C.~A., {\"O}zel, F., \& Psaltis, D.\ 2016, \prc, 93, 32801.

\bibitem[Reichardt et al.(2019)]{Reichardtetal2019} Reichardt T.~A., De Marco O., Iaconi R., Tout C.~A., Price D.~J., 2019, MNRAS, 484, 631 

\bibitem[Quataert \& Shiode(2012)]{QuataertShiode2012} Quataert, E., \& Shiode, J.\ 2012, \mnras, 423, L92

\bibitem[Sabach \& Soker(2014)]{SabachSoker2014} E. Sabach, and N. Soker, \mnras, 439, 954 (2014)

\bibitem[Schreier, \& Soker(2016)]{SchreierSoker2016} Schreier, R., \& Soker, N.\ 2016, Research in Astronomy and Astrophysics, 16, 70

\bibitem[Segev et al.(2019)]{Segevetal2019} Segev, R., Sabach, E., \& Soker, N.\ 2019, arXiv e-prints , arXiv:1904.11331

\bibitem[Soker(2004)]{Soker2004} Soker, N.\ 2004, \na, 9, 399

\bibitem[Soker(2013)]{Soker2013} Soker, N.\ 2013, arXiv:1302.5037

\bibitem[Soker(2016a)]{Soker2016Magnetar} Soker, N.\ 2016a, \na, 47, 88 

\bibitem[Soker(2016b)]{Soker2016Rev} Soker, N.\ 2016b, \nar, 75, 1 

\bibitem[Soker(2017a)]{Soker2017RAA} Soker, N.\ 2017a, Research in Astronomy and Astrophysics, 17, 113

\bibitem[Soker(2017b)]{Soker2017IIb} Soker, N.\ 2017, \mnras, 470, L102.

\bibitem[Soker(2019)]{Soker2019SCPMA} Soker, N.\ 2019, Science China Physics, Mechanics, and Astronomy, 62, 119501 

\bibitem[Soker \& Gilkis(2017a)]{SokerGilkis2017Magnetar} Soker, N., \& Gilkis, A.\ 2017a, \apj, 851, 95

\bibitem[Soker \& Gilkis(2017b)]{SokerGilkis2017} Soker, N., \& Gilkis, A.\ 2017b, \mnras, 464, 3249

\bibitem[Soker \& Gilkis(2018)]{SokerGilkis2018iPTF14hls} Soker, N., \& Gilkis, A.\ 2018, \mnras, 475, 1198 


\bibitem[Sollerman et al.(2019)]{Sollermanetal2019} Sollerman, J., Taddia, F., Arcavi, I., et al.\ 2019, \aap, 621, A30

\bibitem[Thomas et al.(2013)]{Thomasetal2013} Thomas, J.~D., Witt, A.~N., Aufdenberg, J.~P., Bjorkman, J. E., Dahlstrom, J. A., Hobbs, L. M., \& York, D. G.\ 2013, \mnras, 430, 1230

\bibitem[Van Winckel(2017a)]{VanWinckel2017a} Van Winckel, H.\ 2017a, in Miroshnichenko A., Zharikov S., Korcakova, D., Wolf M., eds, ASP Conf. Ser. Vol. 508, The B[e] Phenomenon: Forty Years of Studies. Astron. Soc. Pac., San Francisco, p. 197

\bibitem[Van Winckel(2017b)]{VanWinckel2017b} Van Winckel, H.\ 2017b, in Planetary Nebulae: Multi-Wavelength Probes of Stellar and Galactic Evolution, Proceedings IAU Symposium No. 323,  eds. X. Liu, L. Stanghellini, and A. Karakas A.C., in press  

\bibitem[Vigna-G{\'o}mez et al.(2019)]{VignaGomezetal2019} Vigna-G{\'o}mez, A., Justham, S., Mandel, I., de Mink, S.~E., \& Podsiadlowski, P.\ 2019, arXiv:1903.02135 


\bibitem[Wang et al.(2018)]{Wangetal2018} Wang, L.~J., Wang, X.~F., Wang, S.~Q., et al.\ 2018, \apj, 865, 95

\bibitem[Wheeler et al.(2015)]{Wheeleretal2015} Wheeler, J.~C., Kagan, D., \& Chatzopoulos, E.\ 2015, \apj, 799, 85 

\bibitem[Woosley(2018)]{Woosley2018} Woosley, S.~E.\ 2018, \apj, 863, 105.


\bibitem[Witt et al.(2009)]{Wittetal2009} Witt, A.~N., Vijh, U.~P., Hobbs, L.~M., Aufdenberg, J. P., Thorburn, J. A., \& York, D. G.\ 2009, \apj, 693, 1946

\bibitem[Zilberman et al.(2018)]{Zilbermanetal2018} Zilberman, N., Gilkis, A., \& Soker, N.\ 2018, \mnras, 474, 1194 


\end{thebibliography}
\end{document}